\documentclass[12pt]{article}
\usepackage{a4wide,epsfig}
\newcommand{\ice}[1]{\relax}

\renewcommand{\thefootnote}{\fnsymbol{footnote}}

\newcommand{\gsim}{\;\rlap{\lower 3.5 pt \hbox{$\mathchar \sim$}} \raise 1pt
 \hbox {$>$}\;}
\newcommand{\lsim}{\;\rlap{\lower 3.5 pt \hbox{$\mathchar \sim$}} \raise 1pt
 \hbox {$<$}\;}

\newcommand{\Lb}{L_{\tilde\omega}}

\begin{document}    

\title{\vskip-3cm{\baselineskip14pt
\centerline{\normalsize\hfill DESY 01--112}
\centerline{\normalsize\hfill hep-ph/0108110}
\centerline{\normalsize\hfill August 2001}
}
\vskip.7cm
Heavy-Light Meson Decay Constant from QCD  Sum Rules
in Three-Loop Approximation}

\author{
{A.A. Penin}\thanks{Permanent address:
Institute for Nuclear Research, Russian Academy of Sciences,
60th October Anniversary Prospect 7a, Moscow 117312, Russia.}
\,\,
and
{M. Steinhauser}
  \\[3em]
  {\normalsize II. Institut f\"ur Theoretische Physik,}\\ 
  {\normalsize Universit\"at Hamburg, D-22761 Hamburg, Germany}
}
\date{}
\maketitle

\begin{abstract}
\noindent
In this paper we compute the  decay constant of the  
pseudo-scalar heavy-light
mesons in the heavy quark effective theory framework of QCD sum rules.  
In our analysis  we
include the recently evaluated three-loop result of order
$\alpha_s^2$ for the heavy-light  current correlator. 
The value of the bottom quark mass, which essentially
limits the accuracy of the sum rules for $B$ meson, is extracted
from the nonrelativistic sum rules for $\Upsilon$ resonances
in  the next-to-next-to-leading approximation. 
We find stability 
of our result with respect to all types of corrections
and the specific form of the sum rule which  
reduces the uncertainty. Our results $f_B=206\pm 20$~MeV 
and $f_D=195\pm 20$~MeV 
for the $B$  and $D$ meson decay constants 
are in impressive agreement with recent lattice calculations. 

\vspace{.2cm}


\end{abstract}


\renewcommand{\thefootnote}{\arabic{footnote}}
\setcounter{footnote}{0}


\section{Introduction}

The decay constant of a pseudo-scalar meson with one heavy and one
light quark constitutes a hadronic quantity which is of primary
phenomenological interest. 
It covers  the strength of  the leptonic
weak decays of  $B$ and $D$ mesons and 
enters as an input quantity into  the analysis of 
the nonleptonic  $B$ and  $D$ meson  decays and    
the $B$--$\bar{B}$ mixing process. 
The latter is of special interest
since it provides a direct source of the information on the
Cabibbo-Kobayashi-Maskawa matrix elements involving the top quark.
Still no experimental information of sufficient
accuracy is  available and the theoretical study of the
decay constant is mandatory.

The first quantitative evaluation of the $B$ and $D$ meson decay
constants $f_B$ and $f_D$
using radiative corrections was performed in~\cite{Aliev:1983ra}
where the QCD sum rules proposed in~\cite{NSVZ,ShiVaiZak79} have been used
together with the perturbative two-loop results  
of ${\cal O}(\alpha_s)$ \cite{Bro81}.
More refined evaluations
followed~\cite{Dominguez:1987ea,Narison:1987qc,Reinders:1988vz,
Colangelo:1991ug,Dom93} 
where it was realized that the accuracy of the decay constants is
significantly limited by the uncertainty of the bottom quark mass.
In \cite{BroGro92,Bagan92,Neu92} the heavy quark effective theory
(HQET) \cite{HQET,HQETrep} has been used to resum the leading 
and next-to-leading logarithms of the heavy quark mass. 
After the renormalization group improvement the two-loop  corrections 
were found to be huge which might be a signal of the importance of the
higher order contributions and questions the reliable determination 
of the decay constants.

The only method to compute hadronic matrix elements 
which is
entirely based on the first principles of QCD are probably the
lattice gauge theory simulations. 
For the $B$ meson decay constant the value
\begin{eqnarray}
  f_B^{\rm lat}&=&200\pm 30~{\rm MeV}
  \,,
\label{latb}
\end{eqnarray}
is cited in the review article~\cite{Bernard:2001ki}  
as an average over different calculations performed on the lattice.
It agrees well with a recent evaluation of $f_B$~\cite{AliKhan:2001jg}
where the result $f_B=204(8)(29)(+44)$~MeV is given.
Here, the first error is statistical, the second due to the
discretization and the third one includes the uncertainty from the
lattice scale.
The numbers obtained for $f_B$ 
from lattice calculations are in reasonable agreement with
the ones obtained from sum rules.
An averaged value for the latter can be found in~\cite{Harrison:1998yr}
\begin{equation}
f_B^{\rm s.r.}=178\pm 42~{\rm MeV}\,.
\label{srb}
\end{equation}
For the $D$ meson decay constant the situation is similar. The 
lattice result~\cite{Bernard:2001ki,AliKhan:2001eg}
\begin{eqnarray}
  f_D^{\rm lat}&=&225\pm 30~{\rm MeV}
  \,,
\label{latd}
\end{eqnarray}
again overshoots the sum-rule estimate
\begin{eqnarray}
  f_D^{\rm s.r.}&=&188\pm 48~{\rm MeV}
  \,,
\label{srd}
\end{eqnarray}
given in Ref.~\cite{Harrison:1998yr}.
In spite of the essential progress in the lattice calculations
their uncertainty is still rather large and 
it is too early to rely solely  on the results of this approach.   
On the other hand, the sum-rule analysis can be essentially 
improved to reach an accuracy which is comparable or even better
than the current accuracy of the lattice calculations and provides
an independent cross check of the latter.
In this paper we perform  the analysis of the 
sum rules for the pseudo-scalar
heavy-light  meson decay constant and improve the
existing calculations with respect to several points which we 
summarize  in the following:\\
(i) 
We include the three-loop perturbative corrections of
${\cal O}(\alpha_s^2)$ which recently became
available~\cite{Chetyrkin:2001mq,CheSte01}. \\
(ii) 
The bottom quark mass, which constitutes a crucial input
for $f_B$, is determined 
using the approach suggested in~\cite{PenPiv1,PenPiv}
in the context of semileptonic decays of the bottom quark.
The latter is also quite sensitive to the pole mass $m_b$.
The basic idea is to replace $m_b$
by the ratio of experimental and theoretical moments
of the nonrelativistic sum rules for $\Upsilon$
resonances. It  is  computed order by order up to the 
next-to-next-to-leading
accuracy  depending on the perturbative input used for the 
evaluation of the sum rules for $f_B$. 
In this context we would like to refer to~\cite{Reinders:1988vz}
where also the $\Upsilon$-resonance sum rules have been used, however, only 
up to order $\alpha_s$.

The charm quark pole mass $m_c$,
necessary for the calculation of $f_D$, is then extracted from the 
HQET relation between $m_b$ and $m_c$. \\ 
(iii) For the evaluation of the decay constant $f_B$ we use both 
Laplace and Hilbert sum rules.  We present, for the first time, 
the explicit formulae of the latter in the framework of HQET.
The comparison of the results obtained with both approaches provides us
with an estimate of the intrinsic uncertainty of the method.

We will show that these new ingredients 
improve significantly the reliability of the 
sum rules and finally the prediction for the decay constant, 
in particular for $f_B$.

This paper is organized as follows.
In Sections~\ref{sec:sr}  and~\ref{sec:hqet} 
we introduce the basic features of the sum rules and the HQET formalism, 
respectively.  In Section~\ref{sec:hqetsr} the renormalization group
improved sum rules for the decay constant are given in the three-loop
approximation. The problem of the heavy quark mass determination
is discussed in Section~\ref{sec:mq}. 
In Section~\ref{sec:num}, we present the numerical 
analysis and Section~\ref{sec:con} contains our conclusions.


\section{\label{sec:sr}Sum rules}
The decay constant of a pseudo-scalar meson $P$ consisting of
a heavy ($Q$) and a light quark ($q$) is defined through the matrix
element
\begin{eqnarray}
\langle 0| j_\mu^a | P(p) \rangle &=& i f_P p_\mu\,,
\label{eq:fPdef}
\end{eqnarray}
where $j_\mu^a=\bar{q}\gamma_\mu\gamma_5 Q$ is the axial-vector
current. In this paper we focus on the $B$  meson
with the bottom quark as the heavy constituent but
present also the analysis of $D$ mesons with charm as heavy flavour. 
We neglect the $SU(3)$ violating effects of the strange quark mass.
The ratio of the decay constants of strange and non-strange
mesons can be reliably computed 
both on the lattice and using QCD sum rules as
an essential part of the uncertainties
cancel~\cite{Harrison:1998yr,Bernard:2001ki,AliKhan:2001jg,AliKhan:2001eg}.

In order to derive the QCD sum rules for the pseudo-scalar heavy-light 
meson decay
constant $f_P$ one considers the  correlator
\begin{eqnarray}
\Psi^a(q^2)&=&
i\int \!{\rm d}x\,e^{iqx} \langle 0|T\partial^\mu j^a_\mu(x)
\partial^\nu j^{a\dagger}_\nu(0)|0 \rangle \,,
\label{Pia}
\end{eqnarray}
which is  related  to the  correlator of the pseudo-scalar currents 
\begin{eqnarray}
\Pi^p(q^2)&=& 
i\int \!{\rm d}x\,e^{iqx} \langle 0|Tj^p(x)j^{p\dagger}(0)|0 \rangle
\end{eqnarray}
by the equation $\partial^\mu j^a_\mu=m_Qj^p$,
where 
$j^p = i (m_Q(\mu)/m_Q) \bar{q}\gamma^5Q$ and
$\bar m_Q(\mu)$ and $m_Q$ are the $\overline{\rm MS}$ and pole 
mass of the heavy quark $Q$, respectively.

Following~\cite{ShiVaiZak79} the correlator $\Psi^a(q^2)$
is evaluated in two ways. In the Euclidean region where $q^2-m_Q^2\ll 0$
it  can be reliably computed in QCD because of the  
asymptotic freedom. The correlator 
gets a perturbative contribution corresponding to the leading
operator in the operator
product expansion (OPE) of the two currents in Eq.~(\ref{Pia}).  
Furthermore there are
power suppressed terms from the vacuum expectation values 
of the higher dimension operators (so-called vacuum condensates)
parameterizing the nonperturbative long-distance effects.
On the other hand, the correlator  can be obtained from the dispersion
integral over the physical states.
In the standard analysis  only the ground state meson is supposed to give a
delta-function contribution to the spectral function.
Assuming local quark-hadron duality
the contribution from the
higher resonances is modeled by the perturbative continuum
starting at some threshold $s_c$ which brings 
an intrinsic uncertainty to the approach.
Finally one arrives at the following  equation
\begin{eqnarray}
\Psi^a_{\rm pt}(q^2) + \Psi^a_{\rm npt}(q^2)&=&
\frac{f_P^2M_P^4}{M_P^2-q^2}+\frac{1}{\pi}\int_{s_c}^\infty \!{\rm d}s\,
\frac{\mbox{Im}\left[\Psi^a_{\rm pt}(s)\right]}{s-q^2}
+\mbox{subtractions}\,,
\label{eq:fP1}
\end{eqnarray}
where $M_P$ is the  meson mass. $\Psi^a_{\rm pt}$ and   
$\Psi^a_{\rm npt}$ are the perturbative and nonperturbative
QCD contributions,  respectively. 
The subtractions needed for the dispersion
integral are not specified explicitly as they will drop out 
in the following. 
To suppress the contribution from higher resonances and to
reduce the uncertainty one can 
perform a Borel transformation of  Eq.~(\ref{eq:fP1}) 
\begin{eqnarray}
\left.{1\over (n-1)!}
\left(-q^2{{\rm d}\over {\rm d}q^2}\right)^n\Psi^a(q^2)
\right|_{\rlap{\lower 5 pt \hbox{\tiny $-q^2/n=M^2$} \raise 4pt  
\hbox{\hspace{-16mm}\tiny $-q^2,n\to\infty$}}}
\hspace{12mm}&=&{1\over\pi}\int_{m_Q^2}^\infty\!{\rm d}s\,
{e^{-s/M^2}\over M^2}{\rm Im}[\Psi^a(s)]\,,
\label{Borel}
\end{eqnarray}
where $M$ is the Borel parameter,
and arrive at so-called Laplace sum rules.
Alternatively, it is possible to consider moments of Eq.~(\ref{eq:fP1})  
\begin{eqnarray}
\left.{1\over n!}\left({{\rm d}\over {\rm d}q^2}\right)^n
\Psi^a(q^2)\right|_{q^2=0}&=&
{1\over\pi}\int_{m_Q^2}^\infty\!{\rm d}s\,{{\rm Im}[\Psi^a(s)]\over s^{n+1}}\,,
\label{mom}
\end{eqnarray}
which leads to  Hilbert sum rules.
To estimate the intrinsic uncertainty of the method we will 
follow both options.
In the above sum rules  the weight functions
cut off the  dispersion integral at a typical hadronic scale
much less than the heavy quark mass so that it is saturated
by the near threshold region where   the heavy quark
is nonrelativistic.  Laplace sum rules are particularly relevant 
for the nonrelativistic HQET analysis because of
the heavy quark mass independent exponential suppression of 
the relativistic momentum region. The situation
is more tricky in the case of Hilbert sum rules 
as is discussed below.


\section{\label{sec:hqet}Heavy quark effective theory}

Systematic description of the  heavy quark nonrelativistic dynamics
and consistent separation of the relativistic effects can be done within
HQET. Let us discuss this issue in more detail. The perturbative
heavy-light quark system involves two dynamical scales: 
the hard scale given by the heavy quark mass and the soft scale 
given by the off-threshold energy 
\begin{eqnarray}
  \tilde\omega&=&\frac{q^2-m_Q^2}{m_Q}
  \,.
\end{eqnarray}
By integrating out the relativistic 
hard modes with the off-shell momentum 
of order $m_Q$ one arrives at HQET which includes $\tilde\omega$ as 
dynamical scale.
The effect of the hard modes is accumulated in the 
Wilson (matching) coefficients
leading to an expansion in $\alpha_s$ along with relativistic
corrections and contributions from higher dimensional operators 
leading to an
expansion in $1/m_Q$. In the hadronic matrix elements 
the latter is converted to an expansion in the dimensionless parameter
$\bar\Lambda/m_Q$ where $\bar\Lambda\approx M_P-m_Q$  describes
the nonperturbative long-distance effects
and remains finite as  $m_Q\to\infty$. 
In the process of scale separation spurious divergences appear
at the intermediate steps which result in the anomalous dimensions of
the effective theory operators and lead to the corrections involving
the large logarithms of the form $\ln(m_Q/\tilde\omega)$. 
These logarithmic corrections 
can be resummed  by solving the effective theory renormalization 
group  equations.

For this purpose
let us  consider the effective theory realization
of the axial-vector current. 
The corresponding connection
between the QCD operator and its HQET counterpart
is given by
\begin{eqnarray}
j^a_\mu &=& C_a(m_Q) \tilde{C}(m_Q) \tilde{j}_\mu^\prime(m_Q)
+{\cal O}\left(1/m_Q\right)\,.
\label{eq:dec}
\end{eqnarray} 
The matching coefficients have been computed in~\cite{BroGro95,Gro98}
up to order $\alpha_s^2$. In the $\overline{\rm MS}$ subtraction scheme
they read
\begin{eqnarray}
C_a(M_Q) &=& 1-\frac{\alpha_s^{(n_f)}(m_Q)}{\pi} \frac{2}{3}
+\left(\frac{\alpha_s^{(n_f)}(m_Q)}{\pi}\right)^2
\Bigg[-\frac{683}{576}-\frac{17\pi^2}{72}-\frac{\pi^2}{18}\ln2
-\frac{11}{36}\zeta(3)
\nonumber\\
&&\mbox{}+n_l\left(\frac{47}{288}+\frac{\pi^2}{36}\right)\Bigg]\,,
\nonumber\\
\tilde{C}(m_Q) &=& 1+\frac{89}{864}
\left(\frac{\alpha_s^{(n_l)}(m_Q)}{\pi}\right)^2\,,
\end{eqnarray}
where $n_l$ is the number of light flavours
and  $\zeta(3)=1.2020569\ldots$ 
is the Riemann $\zeta$-function.
The renormalization group equation which governs  the  evolution
of $\tilde{j}^\prime_\mu(\mu)$ is of the form
\begin{eqnarray}
  \mu^2\frac{{\rm d}}{{\rm d}\mu^2} \tilde{j}_\mu^\prime(\mu) &=& 
  \tilde{\gamma}^\prime \tilde{j}_\mu^\prime(\mu)\,,
  \label{eq:rgeq}
\end{eqnarray}
where the anomalous dimension is known up to two loops~\cite{JiMus91}
\begin{eqnarray}
\tilde{\gamma}^\prime &=&\gamma_0 \frac{\alpha_s^{(n_l)}}{\pi}
+\gamma_1 \left(\frac{\alpha_s^{(n_l)}}{\pi}\right)^2+{\cal O}(\alpha_s^3)\,,
\nonumber\\
\gamma_0 &=& \frac{1}{2}\,,\quad
\gamma_1 \,\,=\,\, \frac{127}{144} + \frac{7\pi^2}{108} - \frac{5}{72} n_l\,.
\label{eq:gam}
\end{eqnarray}
The solution of Eq.~(\ref{eq:rgeq}) reads
\begin{eqnarray}
\tilde{j}^\prime_\mu(\mu)&=&\sqrt{X(m_Q)\over X(\mu)}
\tilde{j}^\prime_\mu(m_Q)\,,
\end{eqnarray}
where 
\begin{eqnarray}
X(\mu) &=& \left(\alpha_s^{(n_l)}(\mu)\right)^{2\gamma_0/\beta_0}
\left[ 1+\left(\frac{\gamma_1}{\beta_0}
-\frac{\gamma_0\beta_1}{\beta_0^2}\right)\frac{\alpha_s^{(n_l)}(\mu)}{\pi} 
+ {\cal O}(\alpha_s^2)             
\right]^2
\,,
\end{eqnarray}
and the first two coefficients of the $\beta$-function are given by
\begin{eqnarray}
\beta_0 \,\,=\,\, \frac{11}{4}-\frac{1}{6}n_l\,,
&& \beta_1 \,\,=\,\, \frac{51}{8}-\frac{19}{24}n_l\,.
\end{eqnarray}
Thus in the nonrelativistic region $\tilde\omega\ll m_Q$
we have the following representation of the  perturbative
part of the correlator~(\ref{Pia})
\begin{eqnarray}
  \mbox{Im}\left[ \Psi^a_{\rm pt}(q^2) \right]
  &=&
  \left(C_a(m_Q)\tilde{C}(m_Q)\right)^2
  \frac{X(\mu)}{X(m_Q)}m_Q^2 
  \mbox{Im}\left[\tilde{\Pi}_{\rm pt}(\tilde\omega)\right] + 
  {\cal O}\left(1/m_Q\right)\,,
\end{eqnarray}
where the universal
HQET current correlator $\tilde{\Pi}_{\rm pt}(\tilde\omega)$
does not depend on $m_Q$ and the spin and parity of the currents. 
Its imaginary part is known up to the three-loop approximation
of ${\cal O}(\alpha_s^2)$ \cite{CheSte01}
\begin{eqnarray}
{\rm Im}[\tilde{\Pi}_{\rm pt}(\tilde\omega)] &=& \frac{3\tilde\omega^2}{8\pi}
\Bigg\{1+\frac{\alpha_s^{(n_l)}(\mu)}{\pi}\left[{17\over 3}+{4\pi^2\over 9}
+\Lb\right]+\left(\frac{\alpha_s^{(n_l)}(\mu)}{\pi}\right)^2
\Bigg[99(15)
\nonumber\\
&&\mbox{}+\left({1657\over 72}+{97\pi^2\over 54}\right)\Lb    
+{15\over 8}\Lb^2+n_l\Bigg(-3.6(4)  +\left(-{13\over12}
-{2\pi^2\over 27}\right)\Lb 
\nonumber\\
&&\mbox{}
-{1\over 12}\Lb^2\Bigg)\Bigg]\Bigg\}\,,
\label{eq:rtilfin} 
\end{eqnarray}
where $\Lb=\ln(\mu^2/\tilde\omega^2)$. The uncertainty in the
non-logarithmic three-loop terms results from the semi-numerical
method used in Ref.~\cite{CheSte01}.
For $\mu\approx\tilde\omega$ the 
HQET correlator does not include large logarithms.
They are all contained in the factor $X(\mu)/X(m_Q)$ which
sums up the leading and
next-to-leading logarithms of the form $\alpha_s^n\ln^n(m_Q/\tilde\omega)$
and  $\alpha_s^{n+1}\ln^n(m_Q/\tilde\omega)$.

On the phenomenological side the HQET  decay constant  is
defined through the matrix element of the HQET current 
\begin{eqnarray}
\langle 0| \tilde{j}_\mu (\mu)|\tilde{P}(v) \rangle &=& 
{i\over \sqrt{2}}\tilde{f}_P(\mu) v_\mu\,,
\end{eqnarray}
where $|\tilde{P}(v) \rangle $ is the
nonrelativistic, {\em i.e.} quantum mechanical, meson state
with velocity $v$. It 
is connected to $f_P$ via the relation
\begin{eqnarray}
f_P \sqrt{M_P} &=& C_a(m_Q)\tilde{C}(m_Q)\tilde{f}_{P}(m_Q)
+{\cal O}\left(1/m_Q\right)\,.
\end{eqnarray}
By using the renormalization group  property of the HQET current
it is convenient to introduce the renormalization group invariant 
quantity 
\begin{eqnarray}
  \tilde{f}_P^r &=&\sqrt{X(\mu)}\tilde{f}_P(\mu)\,,
\end{eqnarray}
which is a universal low-energy parameter of
strong interactions.


\section{\label{sec:hqetsr}The heavy quark effective theory sum rules}

Now we are in the position to write down the renormalization group improved
sum rules for the heavy-light pseudo-scalar meson decay constant.

\subsection{Laplace sum rules}

Let us start with the  sum rules in the infinite heavy quark mass limit
$m_Q\to\infty$.
Transforming Eq.~(\ref{eq:fP1}) to HQET 
and neglecting the mass suppressed terms we obtain
\begin{eqnarray}
\left(\tilde{f}_P^r\right)^2 &\!\!\!=\!\!& {\rm e}^{\Delta/T}
\Bigg\{ 
X(\mu)
\Bigg[
\frac{T}{\pi}\int_0^{\tilde\omega_c/T} 
\!\!{\rm d}z\, e^{-z} {\rm Im} [\tilde{\Pi}_{\rm pt}(zT)]
-
\langle\bar{q}q\rangle(\mu)
\left[1+2\frac{\alpha_s}{\pi}
\right]
\Bigg]
+
X(T)
\frac{m_0^2\langle\bar{q}q\rangle}{4T^2}
\Bigg\},
\nonumber\\
\label{eq:fpren}
\end{eqnarray}
where  $\Delta=(M_P^2-m_Q^2)/m_Q$,
$\tilde\omega_c=(s_c-m_Q^2)/m_Q$,
$m_0^2=
\langle\bar{q}g_sG^{\mu\nu}\sigma_{\mu\nu}q\rangle/\langle\bar{q}q\rangle$
and $m_QT=M^2$.
We keep the operators up to dimension five in the OPE and neglect the  
running of the quark-gluon operator.

The   mass suppressed contribution to the correlator  
can be found by subtracting the
asymptotic HQET result from the full theory expression.   
In this way 
the mass suppressed contribution to the 
physical decay constant 
from Laplace sum rules
is obtained as
\begin{eqnarray}
\delta f_{P}^2 &=& {\rm e}^{\Delta/T}
\frac{1}{M_P}\left(\frac{m_Q}{M_P}\right)^3
\Bigg\{ \frac{T}{\pi}\int_0^{\tilde\omega_c/T} 
\!{\rm d}z\, e^{-z} {\rm Im} [\delta\Pi^p_{\rm pt}](zT)
\nonumber\\
&&\mbox{}+\langle\bar{q}q\rangle(m_Q)
\left[\frac{2\alpha_s}{\pi}\frac{T}{m_Q}\int_0^\infty \!{\rm d}z\, 
\frac{{\rm e}^{-z}}{1+zT/m_Q}
\right]
-\frac{m_0^2\langle\bar{q}q\rangle}{2Tm_Q}
+{\langle\alpha_s G^{\mu\nu}G_{\mu\nu}\rangle\over 12\pi m_Q }
\Bigg\}\,,
\label{massup}
\end{eqnarray}
where
\begin{eqnarray}
\mbox{Im}\,\left[\delta\Pi^p_{\rm pt}(q^2)\right]&=&
\mbox{Im}\,\left[\Pi^p_{\rm pt}(q^2)\right]
-\left(C_a(m_Q)\tilde{C}(m_Q)\right)^2\frac{X(\mu)}{X(m_Q)}
\mbox{Im}\,\left[\tilde{\Pi}_{\rm pt}(\tilde\omega)\right]\,.
\label{delPi}
\end{eqnarray}
The one-loop expression for this function is given by  
\begin{eqnarray}
{\rm Im} [\delta^{(1)}\Pi^p_{\rm pt}(\tilde\omega)] &=&
-\frac{3}{8\pi}{\tilde\omega\over m_Q}{\tilde\omega^2\over 1+\tilde\omega/m_Q}
\,.
\label{supp0}
\end{eqnarray}
The  two-loop  approximation for $\Pi^p_{\rm pt}(q^2)$ 
is known in full QCD in  analytical form \cite{Bro81}.
It determines  the ${\cal O}(\alpha_s)$  part 
of Eq.~(\ref{delPi}) which reads
\begin{eqnarray}
\hspace*{-5mm}
{\rm Im} [\delta^{(2)}\Pi^p_{\rm pt}(\tilde\omega)] &=&
\frac{\alpha_s}{2\pi^2}{\tilde\omega^2\over 1+\tilde\omega/m_Q}
\Bigg[-{\tilde\omega\over m_Q}
\left({13\over 4}+{\pi^2\over 3}
+{3\over 2}\ln\left({m_Q\over \tilde\omega}\right)\right)
 +F\left({\tilde\omega\over m_Q}\right)\Bigg]\,,
\label{supp1}
\end{eqnarray}
where 
\begin{eqnarray}
F(x)&=&2{\rm Li}_2(-x)+
\ln(x)\ln(1+x)-{x\over 1+x}\ln(x)+{1+x\over x}\ln(x+1)-1
\nonumber\\
&=&-{3\over 2}x
+\left({1\over 3}+{1\over 2}\ln(x)\right)x^2+{\cal O}(x^3)\,,
\end{eqnarray}
with ${\rm Li}_2(z)$ being the dilogarithmic function. 
The perturbative mass 
suppressed contribution of ${\cal O}(\alpha_s^2)$ 
can be obtained from the numerical three-loop
result of \cite{Chetyrkin:2001mq,CheSte01}
which is available under the URL
\verb|http://www-ttp.physik.uni-karlsruhe.de/Progdata/|
\verb|ttp00-25|.
Thus we get the final expression for the decay constant 
\begin{eqnarray}
f_P^2 &=&   \left(\frac{m_Q}{M_P}\right)^3
\frac{\left(C_a(m_Q)\tilde{C}(m_Q)\right)^2}{X(m_Q)}
{\left(\tilde{f}_P^r\right)^2\over M_P}
+\delta f_{P}^2\,. 
\label{eq:fPfin}
\end{eqnarray}
The first term of this equation includes the leading 
HQET contribution up to ${\cal O}(\alpha_s^2)$
and the resummed leading and next-to-leading logarithms
of the heavy quark mass. The second term represents
all heavy quark mass suppressed terms up to order 
$\alpha_s^2$. 
 
\subsection{Hilbert sum rules}

The use of HQET for Hilbert sum rules is rather  subtle
as they do not have a proper 
infinite heavy quark mass  limit. Indeed, the naive
limit $m_Q\to\infty$ leads to the decay constant
\begin{eqnarray}
\left(\tilde{f}^r_{P}\right)^2 &=&{1\over 8\pi^2}\tilde\omega_c^3-
\langle\bar q q\rangle +{\cal O}(\alpha_s) \,,
\label{Hilstat}
\end{eqnarray}
which, in contrast to Eq.~(\ref{eq:fpren}), does not contain a
dynamical constraint on the parameter $\tilde\omega_c$.
Therefore one cannot use them 
to study the HQET decay constant. Nevertheless it is possible to 
apply  Hilbert sum rules 
for the calculation of the physical decay constant and,
furthermore, use HQET for the analysis. Indeed, if we keep
the factor $1/s^{n+1}=1/(m_Q^{2n+2}(1+\tilde\omega/m_Q)^{n+1})$ 
unexpanded in $1/m_Q$,  it is straightforward  to 
obtain the proper scaling 
\begin{eqnarray}  
\tilde\omega_c&=&\frac{4}{3}\Delta+ {\cal O}\left(\alpha_s,1/m_Q\right)\,,
\label{scale}
\end{eqnarray}
where $\Delta$ is defined after Eq.~(\ref{eq:fpren}),
from the ratio of two arbitrary moments. Note that  Eq.~(\ref{scale})
is obtained from the purely perturbative
correlator and the quark condensate  contribution  is neglected. 
This allows for the nonrelativistic treatment
of the heavy quark  in $\Psi^a(s)$.
Moreover, the  dispersion integral is saturated by the region
$\tilde\omega<m_Q/n$
and thus for $n>m_Q/\tilde\omega_c$ the result  is not sensitive to
$\tilde\omega_c$ in contrast to Eq.~(\ref{Hilstat}). 
In this way we obtain
the renormalization group improved Hilbert sum rules 
of the following form 
\begin{eqnarray}
f_P^2 &=& \frac{M_P^{2n-2}}{m_Q^{2n-1}}\Bigg\{ 
\frac{\tilde\omega_c}{\pi}\int_0^1 
\!{{\rm d}z \over  (1+z\tilde\omega_c/m_Q)^{n+1}}  
\Bigg(\left(C_a(m_Q)\tilde{C}(m_Q)\right)^2{X(\mu)
\over X(m_Q)}{\rm Im} [\tilde{\Pi}_{\rm pt}(z\tilde\omega_c)]+
\nonumber\\  
&& {\rm Im} [\delta\Pi^p_{\rm pt}(z\tilde\omega_c)]\Bigg)
-\langle\bar{q}q\rangle(m_Q)\Bigg[ 1 +  {2\over 3}\frac{\alpha_s}{\pi}
\left(1-3\frac{\tilde\omega_c}{m_Q}\int_0^\infty \!
\frac{{\rm d}z}{(1+z\tilde\omega_c/m_Q)^{n+2}}\right)
\Bigg]
\nonumber\\
&&+{n(n+1)\over 8}\frac{m_0^2\langle\bar{q}q\rangle}{m_Q^2}
+{\langle\alpha_s G^{\mu\nu}G_{\mu\nu}\rangle\over 12\pi m_Q } \Bigg\}\,.
\label{Hil}
\end{eqnarray}


\section{\label{sec:mq}The heavy quark masses}

Before turning to the numerical analysis we want to
discuss the determination of the  heavy quark mass
which is an input parameter of the sum rules given above.
The result for the physical decay constant ~(\ref{eq:fPfin})
is rather sensitive to the  heavy quark mass value. 
Therefore  $m_Q$  should
be determined with a great accuracy in order to obtain
a reasonable precision for $f_P$. 
The best accuracy of the  bottom quark  mass  
determination   is achieved within  the 
heavy quarkonium sum rules \cite{NSVZ}. The corresponding expression for 
the pole  mass is given by the ratio
\begin{eqnarray}
m_b &=& \left({{\cal M}^{\rm th}_n\over 
\tilde{{\cal M}}_n^{\rm exp}}\right)^{1\over 2n}\,.
\label{mb}
\end{eqnarray}
Here the  dimensionful   experimental  moments 
\begin{eqnarray}
\tilde{\cal M}_n^{\rm exp} &=& 
9\int_0^\infty\!{\rm d}s\,{R^{\rm exp}(s)\over s^{n+1}}
\,,
\label{expmom0}
\end{eqnarray}
are generated by the normalized cross section of $e^+e^-$ annihilation
$R^{\rm exp}(s) = \sigma(e^+e^-\rightarrow {\rm hadrons}_{\,b\bar b})/
\sigma(e^+e^-\rightarrow \mu^+\mu^-)$.
The  dimensionless theoretical moments  
are defined as follows
\begin{eqnarray}
{\cal M}_n^{\rm th} &=& 
12\pi(4m_b^2)^n\int_0^\infty\!{\rm d}s\,{{\rm Im}[\Pi^v(s)]\over s^{n+1}}\,,
\label{eq:mth}
\end{eqnarray}
where the vector current correlator is defined through
\begin{eqnarray}
\left(q^\mu q^\nu - q^2g^{\mu\nu}\right)\Pi^v(q^2) &=&
i \int\!{\rm d}x\,e^{iqx}\langle 0|Tj^{\mu}(x)j^{\nu}(0)|0\rangle\,,
\end{eqnarray}
with  $j_\mu=\bar b\gamma_\mu b$.
If $n$ is large enough the experimental moments are
saturated by the  $\Upsilon$ resonance contributions
which is known with high precision. For large $n$ the dispersion integral 
in Eq.~(\ref{eq:mth}) is saturated by the 
region near threshold
where the nonrelativistic expansion in the heavy quark velocity 
is applicable and the correlator
can be systematically computed within the 
effective theory of nonrelativistic QCD (NRQCD) \cite{CasLep}. 
The complete result  for the moments including the 
second order corrections  in the strong coupling constant
and heavy quark velocity is  now available 
\cite{PenPiv,KPP,Hoa,MelYel}.    

It is widely believed that 
due to the renormalon contributions
the absolute value of 
the heavy quark pole mass obtained through Eq.~(\ref{mb})
is divergent~\cite{MelYel,BenSig,Hoa1}. 
As a consequence the absolute value of the 
pole mass is plagued with an intrinsic uncertainty of order $\Lambda_{QCD}$.
On the other hand,  $m_b$ is not an observable and has no
immediate physical meaning. Therefore it can safely be removed from 
relations between physical observables. Using    this philosophy 
we  replace $m_b$ in the sum rules for $f_B$
by the fixed order expression of the right-hand side of  Eq.~(\ref{mb}). 
Equivalently,  we 
determine  the  value to the pole mass according to Eq.~(\ref{mb})
only in a given   order of the perturbative expansion 
correlated to the order of the approximation for $f_B$. 
A detailed discussion of the 
sum rules and the
corresponding numerical results can be found in \cite{PenPiv}.
In particular, we use the
next-to-leading order (NLO) result, $m_b=4.68$~GeV, 
for the calculation of $f_B$ to order $\alpha_s$ and
the next-to-next-to-leading order (NNLO) result, $m_b=4.79$~GeV, 
for the ${\cal O}(\alpha_s^2)$  analysis of  $f_B$. 
The accuracy of the numerical value
for the fixed order approximation for  $m_b$ is no longer
restricted by $\Lambda_{QCD}$ since it is not related to 
the divergence of the series in Eq.~(\ref{mb}).
It is mainly due to the dependence of the 
theoretical moments on the normalization
scale of $\alpha_s$ and due to the  dependence on $n$ 
which altogether amounts to $\pm 60$~MeV  for the NNLO
result \cite{PenPiv,KPP}.  
Note  that the direct order-by-order matching of the 
results for $m_b$ and $f_B$ is not obvious due to different
kinds of  resummation adopted in   the sum rules:
the study of the $\Upsilon$-resonance sum rules  
requires the resummation of the singular
Coulomb terms while the above analysis 
of the $B$-meson sum rules   
involves the resummation of the heavy quark mass
logarithms.  
Our matching  of the perturbative series 
is based on the fact that in a given approximation  both the 
$\Upsilon$-resonance sum rules  and 
the $B$-meson sum rules 
include the  perturbative corrections of the same order in $\alpha_s$
to the  threshold behaviour of the heavy-heavy and heavy-light quark 
current correlators, respectively.
Performing the analysis in the described way 
we expect that the large perturbative corrections to $m_b$ cancel 
in the complete expression so that the final
series for $f_B$  in terms of physical
moments~(\ref{expmom0}) is convergent.  
This approach 
turned out to be very efficient for the analysis
of the bottom quark semileptonic decay width \cite{PenPiv1,PenPiv}. 
We will show  that the method works in the case of the 
sum rules for  $f_B$ as well.

For a given value of the  bottom quark mass  
the charm quark mass can be obtained 
from the  HQET constraint  of the form
\begin{eqnarray}
m_b-m_c+{\cal O}(1/m_{b,c})&=&
M_B - M_D \,,
\label{const}
\end{eqnarray}
which results in $m_c=1.37$~GeV for the NNLO value of 
$m_b$. Due to the cancellation between the different 
terms of order $1/m_{b,c}$ (see, for example, \cite{N}) 
this numerical value is valid 
with  ${\cal O}(1/m_{b,c})$ accuracy. The use of the 
relation~(\ref{const}) brings an additional uncertainty to
$m_c$ so that the total uncertainty can be roughly estimated as 
$\pm 100$~MeV.


\section{\label{sec:num}Numerical analysis}

In this section we  present the numerical analysis of the sum rules.
We adopt the same input values for the vacuum condensates
as in~\cite{Harrison:1998yr}:
\begin{eqnarray}
\langle\bar{q}q \rangle\left(1~\mbox{GeV}\right)
&=& -\left(225(25)~\mbox{MeV}\right)^3\,,
\nonumber\\
\langle \alpha_s G^{\mu\nu}G_{\mu\nu} \rangle &=& 0.04(2)~\mbox{GeV}^4\,,
\nonumber\\ m_0^2\left(1~\mbox{GeV}\right) &=& 0.8(2)~\mbox{GeV}^2 \,.
\end{eqnarray}
The strong coupling constant is evaluated with four active flavours
using two-loop accuracy and $\Lambda^{(4)}=296$~MeV.
This value corresponds to $\alpha^{(5)}_s(m_b)=0.210$ obtained
from $\alpha^{(5)}_s(M_Z)=0.1185$ using the four-loop
renormalization group evolution \cite{rundec}. For the meson masses we use 
$M_B=5.2793(7)$~GeV and $M_D=1.8641(10)$~GeV, respectively
\cite{pdb2000}.

\subsection{The decay constant within heavy quark effective theory}

Let us start  with the analysis of the limit $m_Q\to\infty$.
The general philosophy for the determination of the decay constant 
from  Laplace sum rules is as follows~\cite{ShiVaiZak79}:
one has to optimize the upper bound of the duality interval,
$\tilde\omega_c$, in such a way that the
value of $\tilde{f}^r_P$ as computed from Eq.~(\ref{eq:fpren})
is stable against a variation of the Borel parameter $T$.
The latter is varied in the  range where both the 
hadronic and QCD representations of the 
correlator can be computed reliably. On the QCD side
of the sum rules the restriction on $T$
is mainly due to the perturbative contribution
because $T$ is an effective  scale of $\alpha_s$ in Eq.~(\ref{eq:fpren}).
Taking into account the large value of the second order 
nonlogarithmic coefficient in   Eq.~(\ref{eq:rtilfin})
we conclude that $T$ cannot be 
chosen much less then $1.5$~GeV where $\alpha_s(T)/\pi\gsim 0.1$
to ensure the convergence of the perturbative series.
Note that the  power suppressed terms  become dangerous
at essentially lower $T$ and that the above restriction
also provides the convergence of the OPE. On the opposite side
of the sum rules the hadronic representation of
the correlator is reliable only for $T\lsim\tilde\omega_c$
which provides the exponential suppression of the
contributions from higher resonances.

The logarithmic dependence of $\tilde\Pi_{\rm pt}(zT)$ on $T$ 
in Eq.~(\ref{eq:fpren}) is quite important for 
the stability of the sum rules. Therefore it is crucial to use the HQET 
renormalization group to get control over the high
order logarithmic contributions.
The leading and next-to-leading logarithms of $T$ can be
summed up  by setting  $\mu=T$ in the factor $X(\mu)$
and in the correlator $\tilde\Pi_{\rm pt}(zT)$. We adopt this 
prescription in our analysis. However, the normalization 
scale of $\alpha_s$ in the  ${\cal O}(\alpha_s^2)$ part of 
$\tilde\Pi_{\rm pt}(zT)$ is not fixed in our approximation
and the corresponding $\mu$-dependence is not compensated 
by $X(\mu)$. 
We do not  use $T$ as the normalization scale here
when determining the optimal value of $\tilde\omega_c$ 
because the resulting  spurious $T$-dependence leads to 
rather unstable  sum rules.  
If the normalization scale  of  $\alpha_s$
in the ${\cal O}(\alpha_s^2)$ contribution is not correlated to $T$
the result has a rather weak dependence on $\mu$
when varying the latter in the same interval as $T$.

Adopting the central values of the input
parameters and $\mu=2$~GeV, 
we obtain for the universal HQET decay constant 
\begin{eqnarray}
  \tilde{f}_P^r &=& 410~({\rm MeV})^{3/2}\,,
\end{eqnarray}
which constitutes an average
for $2.05~\mbox{GeV}\le\tilde\omega_c\le2.10$~GeV.
For these values the 
highest stability is observed.
This value should be compared with the result
obtained using the order $\alpha_s$
expression of the correlator $\tilde\Pi_{\rm pt}(\tilde\omega)$
which  is  $\tilde{f}_P^r=418$~{\rm MeV}
at the optimal value  $\tilde\omega_c=2.4$~GeV.
One notices that the inclusion of the ${\cal O}(\alpha_s^2)$
contribution leads to a rather small variation of 
$\tilde{f}_P^r$   
though the correction to the correlator itself is quite large
(cf. Eq.~(\ref{eq:rtilfin})).
This can be explained by a considerable compensation of the
large corrections to $\tilde\Pi_{\rm pt}(\tilde\omega)$ and $m_b$ 
(which enters the analysis through $\Delta$) 
and the change of the optimal value $\tilde\omega_c$.

\subsection{$B$ meson decay constant}

\begin{figure}[t]
  \begin{center}
    \begin{tabular}{c}
      \leavevmode
      \epsfxsize=14cm
      \epsffile[50 240 540 580]{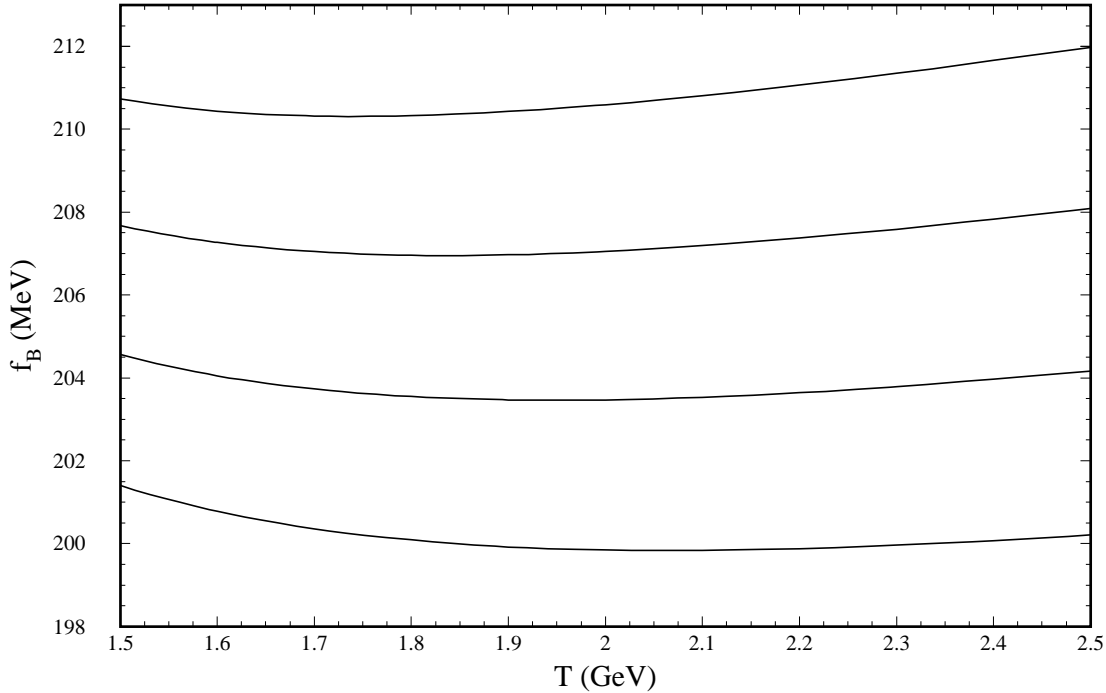}
    \end{tabular}
  \end{center}
  \vspace{-2.em}
  \caption{\label{fig:fB1}The $B$ meson decay constant $f_B$
  as a function of the Borel parameter $T$ of the  
  Laplace sum rules for different values of the 
  threshold parameter $\tilde\omega_c$.
  From top to bottom the curves correspond to 
  $\tilde{\omega}_c=2.35,2.3,2.25$ and $2.2$~GeV.
          }
\end{figure}

Taking into account the mass suppressed contribution to the
Laplace sum rules and performing the analysis along the line
described in the previous section, we obtain
for the $B$ meson decay constant
\begin{eqnarray}
f_B &=& 206~\mbox{MeV}\,,
\label{fB}
\end{eqnarray}
which constitutes an average
for $2.25~\mbox{GeV}\le\tilde\omega_c\le2.3$~GeV.
To illustrate the stability of the sum rules
with respect to the variation of the parameter $T$ 
we plot in Fig.~\ref{fig:fB1} $f_B$  
as a function of $T$ for various values of $\tilde\omega_c$. 
It can be seen that the curves for
$\tilde\omega_c\approx 2.25$~GeV and
$\tilde\omega_c\approx 2.3$~GeV provide the most stable results.
Note that for these values of $\tilde\omega_c$ the function   
$f_B(T)$ has a weakly pronounced minimum around $T\approx 2$~GeV.
At the same time, the use of the two-loop ${\cal O}(\alpha_s)$ approximation
of $\tilde\Pi_{\rm pt}(\tilde\omega)$ along with the NLO value
of $m_b$ gives $f_B=205$~MeV
at the optimal value $\tilde\omega_c=2.75$~GeV.
As in the heavy quark limit  
we observe that due to the compensation of the corrections
the ${\cal O}(\alpha^2_s)$ result 
is practically the same as the ${\cal O}(\alpha_s)$ one.
This fact is a strong argument in favour of our 
treatment of the bottom quark mass. 
The ${\cal O}(\alpha^2_s)$ mass suppressed 
corrections, which are included in Eq.~(\ref{fB}),
reduce the value of $f_B$ by approximately $5$~MeV.
Taking into account 
the fact that the nonperturbative part of the QCD contribution is saturated
by the leading quark condensate, which gives about $10\%$ of the total QCD
contribution, we conclude that
our result is stable with respect to all the types of
corrections to the sum rules.

Note that no  rigorous results concerning  
the high order behaviour of the perturbative series 
both for the $\Upsilon$-resonance and $B$-meson sum rules  
are  available and the absence of sizable higher order  
perturbative corrections to the decay constant within our
approach  cannot be proven strictly  even though it works 
well up to the ${\cal O}(\alpha^2_s)$ approximation. 
However,  our approach provides the convergence  
also in higher orders if the divergence   
of the perturbative series  for the $\Upsilon$-resonance 
and $B$-meson sum rules is related to the use of the  pole mass 
and is thus dominated  by the  renormalon contribution. This is 
because  we effectively remove the pole mass from 
the analysis of the decay constant
in favour of the experimental moments of the $\Upsilon$-resonance
spectral density and operate with the relation between
physical observables  which is free from the  corresponding 
renormalon ambiguity.

Let us next discuss the uncertainty of the result in Eq.~(\ref{fB}).
The error in $m_b$ of $\pm 60$~MeV~\cite{PenPiv1,PenPiv}
results in an uncertainty of $\pm 12$~MeV in $f_B$. 
The variation of the input value of $\alpha_s(M_Z)$
within the experimental error bars $0.1185\pm 0.0020$
\cite{pdb2000} leads to the uncertainty 
interval $201~{\rm MeV}<f_B<213~{\rm MeV}$.
On the other hand, the result is not sensitive to the 
normalization point of $\alpha_s$ in the ${\cal O}(\alpha_s^2)$
contribution
and to the non-logarithmic three-loop coefficient in Eq.~(\ref{eq:rtilfin}).
Note, that the change of the parameters requires every time a new
optimization of $\tilde\omega_c$.

Another source of errors is the intrinsic uncertainty 
of the method due to the approximation of the hadronic
contribution to the dispersion integral~(\ref{eq:fP1}).
A rough estimate of this uncertainty is obtained by the
variation of the upper bound of the duality interval 
around its optimal value. The variation of $\tilde\omega_c$ 
by $\pm 100$~MeV  leads
to $\pm 7$~MeV variation of $f_B$ which can be read off 
Fig.~\ref{fig:fB1}. A larger deviation from the optimal
value leads to the essentially unstable sum rules. 
If we add the errors induced by the uncertainties in $m_b$, $\alpha_s$
and $\tilde{\omega}_c$ discussed so far in quadrature we obtain
$f_B=206\pm16$~MeV.

\begin{figure}[t]
  \begin{center}
    \begin{tabular}{c}
      \leavevmode
      \epsfxsize=14cm
      \epsffile[50 240 540 580]{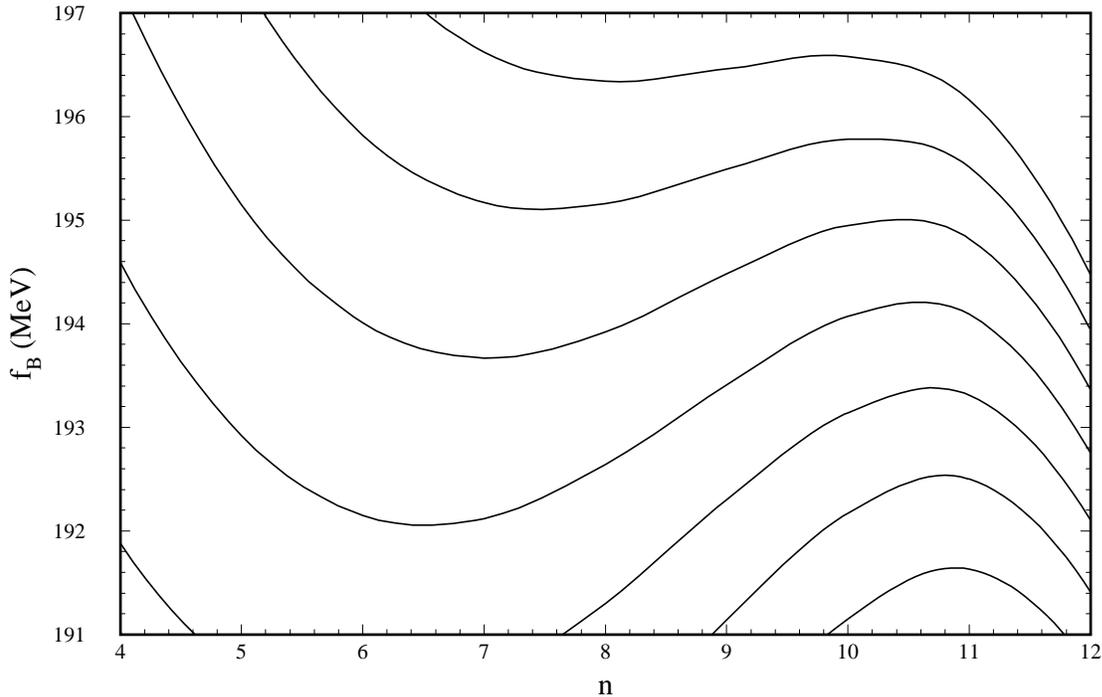}
    \end{tabular}
  \end{center}
  \vspace{-2.em}
  \caption{\label{fig:fB2}The $B$ meson decay constant $f_B$
  as a function of the  moment number
  $n$  of the Hilbert sum rules for different values of the 
  threshold parameter $\tilde\omega_c$.
  From top to bottom the curves correspond to 
  $\tilde{\omega}_c=2.35,2.3,2.25,2.2,2.15,2.1$ and $2.05$~GeV. 
  The value $\mu=2$~GeV has been adopted.
          }
\end{figure}

A more advanced way to estimate the intrinsic uncertainty
of the sum-rule approach is to change the weight function
in the dispersion integral and redo the analysis using 
the sum rules which operate with Hilbert moments of the 
correlator~(\ref{mom}) instead of its Borel transform~(\ref{Borel}). 
The range of  
$n$ relevant for reliable predictions of Hilbert sum rules
is, in fact, quite restricted.
The contribution of the mixed condensate 
grows rapidly with $n$. Thus, requiring  the convergence 
of the OPE sets an upper limit on $n$.  For 
the bottom quark it is $n\approx 12$ where the 
contribution of the  mixed condensate is approximately 
$2/3$ of the leading quark one.  At the same time,
to avoid strong dependence of the result on $\tilde\omega_c$,
one should use $n> m_b/\tilde\omega_c\approx 3$.  
The optimal value of $\tilde\omega_c$ 
can be found by minimizing the dependence of the result
on $n$ in the above interval which results to
$\tilde\omega_c\approx2.2$~GeV. For this value the decay constant 
stays within the interval  $192~{\rm MeV}\lsim f_B\lsim 195~{\rm MeV}$ as $n$ 
varies from 4 to 12  as can be seen in Fig.~\ref{fig:fB2}
where $f_B$ is plotted as a function of $n$.
This result for $f_B$ 
is in good agreement with the value obtained from the Laplace sum rules.

Note, that in addition to  the stability requirement there is
a strong  consistency check on the value of  $\tilde\omega_c$.
The decay constant drops out in the ratio of two 
moments which leads to a relation between $M_B$, $m_b$ and  
$\tilde\omega_c$. Our analysis is consistent 
if the physical value of the meson mass is reproduced  
from this relation 
for some value of the bottom quark mass in the interval
given by the sum rules for the $\Upsilon$ resonances.
For the above value of $\tilde\omega_c$ this requirement
is fulfilled for all $n$ in the allowed interval. 
In fact, the ratio of the $6^{\rm th}$ and $7^{\rm th}$ moment
and the ratio of the 
$10^{\rm th}$ and $11^{\rm th}$ moment imply exactly
the central value $m_b=4.79$. 
Note that for these moments $f_B$ has 
a local extremum as a function of $n$. In other words, if 
the $B$-meson sum rules in the three-loop approximation
are used to determine $m_b$,
the result is in perfect agreement with the NNLO value 
of the bottom quark mass 
from the $\Upsilon$-resonances sum rules. 

We would like to emphasize that the dependence of 
$f_B$ on $m_b$ is completely different for the Laplace
and Hilbert sum rules. 
Thus, by comparing the results of the sum rules
one can also estimate the error due to the uncertainty in $m_b$. 
Furthermore, the prescription 
how the parameter $\tilde\omega_c$ is determined and the 
structure of the condensate contributions are also quite different.
The fact that both approaches
give close results reflects the small intrinsic uncertainty
of the sum-rule method in this particular case and
furthermore strongly supports our treatment of $m_b$.  
Thus the total error originating from the uncertainty in $m_b$
and the approximation of the hadronic spectrum 
can be estimated as $\pm 15$~MeV. The remaining error
is mainly due to the uncertainty in the input values of $\alpha_s(M_Z)$.   
Consequently, as a conservative estimate of the uncertainty of our 
result for $f_B$ we quote $\pm 20$~MeV.
Because of the stronger dependence of the Hilbert sum rules 
on $m_b$ we use them to estimate the error but take
the central value for our final result for $f_B$
from the Laplace sum rules.  

\subsection{$D$ meson decay constant}

Since $m_c$ is not large in comparison to
the scale $\bar\Lambda$ the expansion in $1/m_c$ 
can not provide us with the same accuracy 
as we have for the bottom quark.
By the same reason no resummation
of  the charm quark mass logarithms is necessary
and we can just use the three-loop result for the correlator up to
${\cal O}(\alpha_s^2)$
in the full theory~\cite{Chetyrkin:2001mq,CheSte01}. 
The use of the Laplace sum rules then leads to 
\begin{eqnarray}
f_D &=& 195~\mbox{MeV}\,,
\end{eqnarray}
for the optimal value $\tilde\omega_c=2.35$~GeV.
The Hilbert sum rules are not reliable in this case because 
of the strong dependence on $m_c$ which is known with 
much less relative accuracy than $m_b$. Furthermore,
the Hilbert sum rules suffer from 
large contribution from the higher dimension condensates.
The variation of the input parameters basically leads to
similar variations of $f_D$ and  $f_B$. Due to the
weaker sensitivity of $f_D$  to the charm quark mass the 
additional uncertainty in $m_c$ does not lead to a larger
error in  $f_D$. Thus, assuming the same intrinsic uncertainty of the 
sum rules we obtain the same error bars for the extracted value
of $f_D$.


\section{\label{sec:con}Conclusions}

To summarize, we have computed the  $B$ and $D$ 
meson decay constants within  the QCD sum rules approach.  
Our final results read
\begin{eqnarray}
f_B &=& 206\pm 20~\mbox{MeV}\,,
\label{fBfin}
\\
f_D &=& 195\pm 20~\mbox{MeV}\,.
\label{fDfin}
\end{eqnarray}
For the analysis
we used the tree-loop result for the  heavy-light current correlator.
The large logarithms of the bottom  quark mass have been taken 
into account by means of  the  HQET renormalization group. 
The bottom quark mass  which essentially
limits the accuracy of the sum rules for $f_B$
has been extracted  from the $\Upsilon$-resonance sum rules up to NNLO.
In the case of the $B$ meson the  analysis has been performed by employing  
Laplace and Hilbert sum rules.  
They  have quite a  different structure especially 
as regards the  dependence on the bottom quark mass.  
The fact that the results obtained with these two approaches 
are in a good agreement gives us  confidence in the reliability of the 
sum-rule method applied to the calculation of the decay constants.   
The result also turned out to be quite stable 
with respect to inclusion of the perturbative corrections in $\alpha_s$
and $1/m_b$ and the nonperturbative corrections due to the 
vacuum condensate contributions.
This allowed us to  reduce  the 
uncertainty of  the extracted value of $f_B$ and $f_D$.
The obtained values of both  the $B$ and $D$ meson decay constants 
are consistent with the existing sum-rule results 
(cf. Eqs.~(\ref{srb}) and~(\ref{srd})). 
However, the accuracy of our 
result is   increased in comparison
to the previous  estimates.    
The values in Eqs.~(\ref{fBfin}) and~(\ref{fDfin})  
are in impressive agreement with the
results obtained in lattice calculations 
(cf. Eqs.~(\ref{latb}) and~(\ref{latd})).
Probably no further  improvement of the accuracy
is possible within  the standard  QCD sum-rule framework 
due to the  intrinsic uncertainty of the method.

Our final comment concerns the current experimental status. 
A measurement is only available for the 
$D_s$ meson decay constant where 
the most recent result reads~\cite{Abbiendi:2001nb}
\begin{eqnarray}
  f^{\rm exp}_{D_s} &=& 286\pm 44({\rm stat})\pm 41({\rm syst})~\mbox{MeV}\,.
  \label{fDexp}
\end{eqnarray}
Converting the number in Eq.~(\ref{fDfin}) with the help 
of the lattice result $f_{D_s}/f_D=1.18$~\cite{AliKhan:2001eg}, 
which agrees with the values given in~\cite{Harrison:1998yr},
we obtain 
$f_{D_s}=230$~MeV. This is in reasonable agreement with the 
experimental value of Eq.~(\ref{fDexp}).

\vspace{1cm}
\noindent
{\bf Acknowledgements}
\smallskip

\noindent
We would like to thank K.G. Chetyrkin for the motivation to perfrom
the calculations presented in this paper
and A. Ali for useful comments and discussions.
This work was supported in part by the Deutsche Forschungsgemeinschaft through
Grant No.\ KN~365/1-1, by the Bundesministerium f\"ur Bildung und Forschung
through Grant No.\ 05~HT9GUA~3, and by the European Commission through the
Research Training Network {\it Quantum Chromodynamics and the Deep Structure
of Elementary Particles} under Contract No.\ ERBFMRX-CT98-0194.






\begin{thebibliography}{99}

\def\ap#1#2#3{  {Ann.\ Phys.\ B }{\bf #1} (#2) #3}
\def\app#1#2#3{  {Act.\ Phys.\ Pol.\ B }{\bf #1} (#2) #3}
\def\cpc#1#2#3{  {Comp.\ Phys.\ Commun.\ }{\bf #1} (#2) #3}
\def\cmp#1#2#3{  {Comm.\ Math.\ Phys.\ }{\bf#1} (#2) #3}
\def\epjc#1#2#3{ {Eur.\ Phys.\ J.\ C }{\bf #1} (#2) #3}
\def\fortp#1#2#3{{Fortschr.\ Phys.\ }{\bf#1} (#2) #3}
\def\jcp#1#2#3{  {J.\ Comp.\ Phys.\ }{\bf#1} (#2) #3}
\def\nima#1#2#3{ {Nucl.\ Inst.\ Meth.\ A }{\bf #1} (#2) #3}
\def\npb#1#2#3{  {Nucl.\ Phys.\ B }{\bf #1} (#2) #3}
\def\nca#1#2#3{  {Nuovo Cim.\ A }{\bf #1} (#2) #3}
\def\plb#1#2#3{  {Phys.\ Lett.\ B }{\bf #1} (#2) #3}
\def\prc#1#2#3{  {Phys.\ Reports }{\bf #1} (#2) #3}
\def\prd#1#2#3{  {Phys.\ Rev.\ D }{\bf #1} (#2) #3}
\def\pR#1#2#3{   {Phys.\ Rev.\ }{\bf #1} (#2) #3}
\def\prl#1#2#3{  {Phys.\ Rev.\ Lett.\ }{\bf #1} (#2) #3}
\def\pr#1#2#3{   {Phys.\ Reports }{\bf #1} (#2) #3}
\def\ppnp#1#2#3{ {Prog.\ Part.\ Nucl.\ Phys.\ }{\bf #1} (#2) #3}
\def\sovnp#1#2#3{{Sov.\ J.\ Nucl.\ Phys.\ }{\bf #1} (#2) #3}
\def\tmf#1#2#3{  {Teor.\ Mat.\ Fiz.\ }{\bf #1} (#2) #3}
\def\yadfiz#1#2#3{{Yad.\ Fiz.\ }{\bf #1} (#2) #3}
\def\zpc#1#2#3{  {Z.\ Phys.\ C }{\bf #1} (#2) #3}
\def\ibid#1#2#3{ {ibid.\ }{\bf #1} (#2) #3}


\bibitem{Aliev:1983ra}
T.~M.~Aliev and V.~L.~Eletsky,
Sov.\ J.\ Nucl.\ Phys.\  {\bf 38} (1983) 936
[Yad.\ Fiz.\  {\bf 38} (1983) 1537].

\bibitem{NSVZ} 
V.~A.~Novikov {\em et al.}, Phys.\ Rev.\ Lett.\ {\bf 38} (1977) 626,
(E) ibid. {\bf 38} (1977) 791;
Phys.\ Rep.\ C {\bf 41} (1978) 1.
   
\bibitem{ShiVaiZak79}
M.~A.~Shifman, A.~I.~Vainshtein, and V.~I.~Zakharov,
\npb{147}{1979}{385}; \npb{147}{1979}{448}.

\bibitem{Bro81}
D.~J.~Broadhurst, \plb{101}{1981}{423}.

\bibitem{Dominguez:1987ea}
C.~A.~Dominguez and N.~Paver,
Phys.\ Lett.\  B {\bf 197} (1987) 423,
[Erratum-ibid.\ B {\bf 199} (1987) 596].

\bibitem{Narison:1987qc}
S.~Narison,
Phys.\ Lett.\ B {\bf 198} (1987) 104;
Nucl.\ Phys.\ Proc.\ Suppl.\  {\bf 74} (1999) 304.

\bibitem{Reinders:1988vz}
L.~J.~Reinders,
Phys.\ Rev.\ D {\bf 38} (1988) 947.

\bibitem{Colangelo:1991ug}
P.~Colangelo, G.~Nardulli, A.~A.~Ovchinnikov, and N.~Paver,
Phys.\ Lett.\ B {\bf 269} (1991) 201.

\bibitem{Dom93}
C.~A. Dominguez, in {\it Proceedings of the Third Workshop on the
Tau-Charm Factory}, Marbella, Spain, 1-6 June (1993), Ed. J. Kirkby
and R. Kirkby, Editions Fronti\'eres, p. 357.

\bibitem{BroGro92}
D.~J.~Broadhurst and A.~G.~Grozin, \plb{274}{1992}{421}.

\bibitem{Bagan92}
E.~Bagan,  P.~Ball, V.~M.~Braun, and H.~G.~Dosch, 
\plb{278}{1992}{457}.

\bibitem{Neu92}
M.~Neubert, \prd{45}{1992}{2451}.

\bibitem{HQET}  
E.~Eichten, B.~Hill, Phys.\ Lett.\ B {\bf 234} (1990) 511;\\
H.~Georgi,  Phys.\ Lett.\  B {\bf 240} (1990) 447.

\bibitem{HQETrep} 
M.~Neubert,  Phys.\ Rept.\ {\bf 245} (1994) 259.
 
\bibitem{Bernard:2001ki}
C.~Bernard,
Nucl.\ Phys.\ Proc.\ Suppl.\  {\bf 94} (2001) 159.

\bibitem{AliKhan:2001jg}
A.~Ali Khan {\it et al.}  [CP-PACS Collaboration],
hep-lat/0103020.


\bibitem{Harrison:1998yr}
P.~F.~Harrison and H.~R.~Quinn  [BABAR Collaboration],
SLAC-R-0504
{\it Papers from Workshop on Physics at an Asymmetric B Factory (BaBar
Collaboration Meeting), Rome, Italy, 11-14 Nov 1996, Princeton, NJ,
17-20 Mar 1997, Orsay, France, 16-19 Jun 1997 and Pasadena, CA, 22-24 Sep
1997}.

\bibitem{AliKhan:2001eg}
A.~Ali Khan {\it et al.}  [CP-PACS Collaboration],
Phys.\ Rev.\ D {\bf 64}, 034505 (2001).

\bibitem{Chetyrkin:2001mq}
K.~G.~Chetyrkin and M.~Steinhauser,
Phys.\ Lett.\ B {\bf 502} (2001) 104.

\bibitem{CheSte01}
K.~G.~Chetyrkin and M.~Steinhauser,
Report Nos.: DESY 01--090, THEP 01/07, TTP01--14 and hep-ph/0108017,
Eur.\ Phys.\ J.\ C (in press).

\bibitem{PenPiv1}
A.~A.~Penin and A.~A.~Pivovarov,
Phys.\ Lett.\ B {\bf 443} (1998) 264.

\bibitem{PenPiv}
A.~A.~Penin and A.~A.~Pivovarov,
Nucl.\ Phys.\ B {\bf 549} (1999) 217.

\bibitem{BroGro95}
D. J. Broadhurst and A. G. Grozin, \prd{52}{1995}{4082}.

\bibitem{Gro98}
A. G. Grozin, \plb{445}{1998}{165}.

\bibitem{JiMus91}
X.~Ji and M.~J.~Musolf, \plb{257}{1991}{409};\\
D.~J.~Broadhurst and A.~G.~Grozin, \plb{267}{1991}{105};\\
V.~Gim\'enez, \npb{375}{1992}{582}.

\bibitem{CasLep} 
W.~E.~Caswell and G.~E.~Lepage, Phys.\ Lett.\   
B {\bf 167} (1986) 437;\\
G.~E.~Lepage {\em et al.},  Phys.\ Rev.\  D {\bf 46} (1992) 4052.

\bibitem{KPP}
J.~H.~K\"uhn, A.~A.~Penin, and A.~A.~Pivovarov,
Nucl.\ Phys.\ B {\bf 534} (1998) 356;\\
A.~A.~Penin and A.~A.~Pivovarov,
Phys.\ Lett.\ B {\bf 435} (1998) 413.

\bibitem{Hoa} A.~H.~Hoang, Phys.\ Rev.\ D {\bf 59} (1999) 014039.

\bibitem{MelYel}
K.~Melnikov and A.~Yelkhovsky,
Phys.\ Rev.\ D {\bf 59} (1999) 114009.

\bibitem{BenSig}
M.~Beneke and A.~Signer,
Phys.\ Lett.\ B {\bf 471} (1999) 233.
         
\bibitem{Hoa1}
A.~H.~Hoang, Phys.\ Rev.\ D {\bf 61} (2000) 034005.

\bibitem{N}
M.~Neubert, {\it ``B Decays And the Heavy Quark Expansion''},\\
in the Second Edition of: {\it Heavy
Flavours}, edited by A.~J. Buras and M. Lindner (World Scientific, Singapore),
hep-ph/9702375.

\bibitem{rundec}
K. G.~Chetyrkin, J. H.~K\"uhn and M.~Steinhauser,
Comput.\ Phys.\ Commun.\  {\bf 133} (2000) 43.

\bibitem{pdb2000}
D.~E.~Groom {\em et al.}, \epjc{15}{2000}{1}.


\bibitem{Abbiendi:2001nb}
G.~Abbiendi {\it et al.}  [OPAL Collaboration],
hep-ex/0103012.


\end{thebibliography}
\end{document}